\documentclass[onecolumn, a4paper,journal,12pt]{IEEEtran}
\usepackage{amsfonts}
\usepackage{graphicx, epsfig,amsmath,amssymb,latexsym,graphics,float}
\usepackage{setspace,subfig}
\usepackage{citesort}

\include{IEEEtran.bib}

\begin{document}

\title{Mixture Modeling based Probabilistic Situation Awareness}
\author{Bin~Liu
\thanks{B. Liu is with Nanjing University of Posts and Telecommunications, Nanjing,
Jiangsu, 210023 China e-mail: bins@ieee.org.}}
\maketitle
\begin{abstract}
The problem of situational awareness (SAW) is investigated from the probabilistic modeling point of view. Taking the situation as a hidden variable, 
we introduce a hidden Markov model (HMM) and an extended state space model (ESSM) to mathematically express the dynamic evolution law of the 
situation and the relationships between the situation and the observable quantities. We use the Gaussian mixture model (GMM) to formulate expert 
knowledge, which is needed in building the HMM and ESSM. We show that the ESSM model is preferable as compared with HMM, since using ESSM, 
we can also get a real time estimate of the pivot variable that connects the situation with the observable quantities. The effectiveness and 
efficiency of both models are tested through a simulated experiment about threat surveillance.
\end{abstract}

\section{Introduction}\label{sec:intro}
Situation awareness (SAW) is a field of study concerned with perception of the environment and is critical to decision-makers in complex, 
dynamic areas from air traffic control, military command and control, ship navigation and aviation to emergency services such as fire fighting 
and policing.

SAW is an important issue in various fields, while there has not yet been a commonly recognized definition for it. To highlight the common themes 
and illustrate the diversity in interpretations, we present some sample definitions here.
\begin{itemize}
\item SAW is ``\emph{adaptive, externally-directed consciousness that has as its products knowledge about a dynamic task environment and directed 
action within that environment}" \cite{smith1995risk}.
\item SAW is ``\emph{principally (though not exclusively) cognitive, enriched by experience}" \cite{hartman1991situational}.
\item ``\emph{In psychological terms, this means (that) SAW involves more than perception or pattern recognition: it doubtless requires use of all 
the higher cognitive functions a person can bring to a task}" \cite{vidulich1994cognitive}.
\item ``$\cdots$\emph{the perception of the elements in the environment within a volume of time and space, the comprehension of their meaning, and 
the projection of their status in the near future}" \cite{endsley1995measurement}.
\end{itemize}
The term SAW has also been recognized as a higher level data fusion mechanism, according to the definition of data fusion given by the Joint Directors 
of Laboratories (JDL) \cite{hinman2002some}, now known as the Data Fusion Information Group (DFIG) \cite{blasch2006level}.

An achieved consensus on SAW is that a SAW system must aggregate state estimates provided by lower level information fusion systems to help users 
understand key aspects of the aggregate situation and project its likely evolution.

Ontologies and Bayesian networks are the common tools to do SAW. Ontologies are used to provide common semantics for expressing information about 
entities and relationships in the SAW domain \cite{kokar2009ontology,carvalho2010prognos}. Probabilistic ontologies are proposed to augment standard 
ontologies with support for uncertainty management \cite{carvalho2010prognos}.
Multi-Entity Bayesian Networks (MEBN), which combine first-order Logic with Bayesian networks, are the logical basis for the uncertainty representation 
in the Probabilistic ontologies of 
SAW \cite{wright2002multi,park2013multi,das2002situation,fischer2013modeling,park2014predictive,fischer2014modeling,naderpour2014intelligent}.
In previous applications of MEBN for SAW, a MEBN Model was usually constructed manually by a domain expert \cite{Carvalho2011}. Manual MEBN modeling 
is a labor-intensive and insufficiently agile process. Therefore a machine learning algorithm was proposed \cite{park2013multi,park2013multi2} to 
learn the structure of the MEBN model. However, such learning based methods are limited to cases when training data are available, while, this 
requirement is seldom satisfied in practice. Further, the learning process is usually complex, not easy to implement, and time-consuming.

In this paper, we propose a novel SAW approach that can get rid of labor-intensive tuning or time-consuming 
learning. Taking the situation as a hidden variable, we introduce a hidden Markov model (HMM) and an extended 
state space model (ESSM) to mathematically express the dynamic evolution law of the situation and the 
relationships between the situation and the observable variables. We then use the Gaussian mixture model 
(GMM) to formulate the the expert knowledge that is necessarily needed in building the HMM and ESSM. 
The inference engine is built based on the stochastic simulation techniques.

It is not our purpose to suggest that the proposed method is superior to any existing method in any general sense. 
Typically it is possible to find problems which are most suitable to any given algorithm at hand 
(and \emph{vice versa}). The goal of this research is to provide an alternative candidate solution to SAW, 
which is robust, easy to implement and does not require labor-intensive tuning or time-consuming model 
learning.

The rest of this paper is organized as follows. In Section II we describe the proposed models and the corresponding 
algorithms. In Section III we present the applications of the proposed approaches in a simulated experiment 
on threat surveillance. In Section IV we conclude the paper.
\section{Mixture Modeling based SAW}\label{sec:model}
In this section, we introduce a hidden Markov model (HMM) and an extended state space model (ESSM) to represent 
the SAW process. In both models, the situation is treated as a hidden variable, and the relationship between 
the situation and the observable quantities is characterized by a likelihood function, which is determined by 
the model structure and expert knowledge. The Gaussian mixture model (GMM) is used to formulate the expert 
knowledge, which is required to define the likelihood function. We begin with an introduction of the HMM based 
formulation of the SAW process. Then we describe the proposed ESSM based approach in detail.
\subsection{Mixture based HMM for SAW}\label{sec:hmm}
\begin{figure*}[htb]
\centerline{\includegraphics[angle=0,height=2.5cm,width=5cm]{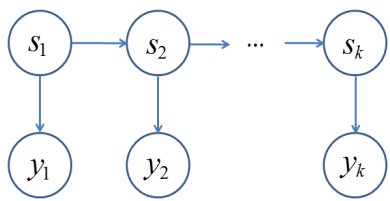}}
\caption{Hidden Markov model for SA}\label{hmm}
\end{figure*}

Here we use the HMM to represent the SA process. A graphical illustration of this model is shown in 
Fig.\ref{hmm}, where $k$ denotes the discrete time step, $s$ and $y$ denote the hidden state variable, 
namely the situation, and the sensor measurement, respectively. In this model, $s$ is a discrete variable, 
whose value space is defined to be $\mathcal{S}\triangleq \{Situation~1, Situation~2, \dots, Situation~m\}$, 
where $m\in\mathbb{R}$ representing the total number of situation elements of our interest. Each arrow in 
Fig.\ref{hmm} indicates a dependence. Therefore, as for this model, the sensor measurement $y$ is 
straightforwardly dependent on the situation $s$ of current time step, and the situation $s_k$ is only 
dependent on $s_{k-1}$.

Given the HMM model structure as shown above, our task is to calculate the posterior probability density 
function (pdf) $p(s_k|y_{1:k})$, where $y_{1:k}=\{y_1,y_2,\dots,y_k\}$, $k=1,2,\ldots$. 
Assume that $p(s_1|y_1)$ is known \emph{a priori}, then what we are really concerned with is that, 
given $p(s_{k-1}|y_{1:k-1})$, how to calculate $p(s_{k}|y_{1:k})$, $k=2,3,\ldots.$. Based on Bayes theorem 
and basic probability calculus, we have
\begin{equation}\label{posterior_hmm}
p(s_k|y_{1:k})=\frac{p(s_{k}|y_{1:k-1})p(y_k|s_k)}{p(y_k|y_{1:k-1})},
\end{equation}
where
\begin{equation}
p(y_k|y_{1:k-1})=\sum_{s_k\in\mathcal{S}} p(s_k|y_{1:k-1})p(y_k|s_k),
\end{equation}
and
\begin{equation}\label{predictive_hmm}
p(s_k|y_{1:k-1})=\sum_{s_{k-1}\in\mathcal{S}} p(s_{k-1}|y_{1:k-1})p(s_k|s_{k-1}).
\end{equation}
It is shown that, in order to calculate the posterior $p(s_k|y_{1:k})$, it is required to be able to 
compute the state transition pdf $p(s_k|s_{k-1})$ and the likelihood function $p(y_k|s_k)$. 
Assume that $p(s_k|s_{k-1})$ is known \emph{a priori}, the focus is on the calculation of the 
likelihood, $p(y_k|s_k)$. As the situation has a higher level semantic meaning, the sensor measurement 
may not be dependent on it directly. For example, if the sensor produces noisy bearing observations of 
moving targets monitored within a surveillance region, it would be difficult to build up the straightforward 
tie between these noisy measurements and the situation. In another word, in that case, there is no easy way to 
calculate the likelihood $p(y|s)$. We solve the above problem by introducing another hidden variable $x$, which 
denotes a time-changing state vector including the position and velocity elements of the targets under 
surveillance. Then we could connect the $y$ with $s$ through $x$, and calculate the likelihood as follows
\begin{equation}\label{likelihood_hmm}
p(y_k|s_k)=\int_{\mathcal{X}} p(x_k,y_k|s_k)dx_k=\int p(y_k|x_k)p(x_k|s_k)dx_k,
\end{equation}
where $\mathcal{X}$ denotes the value space of $x_k$. It is shown that, the underlying assumption is that given $x_k$, $y_k$ is independent with $s_k$, which is intuitively reasonable. Under this assumption, to calculate the likelihood $p(y_k|s_k)$, we need to compute $p(x_k|s_k)$, which is determined by the relationship between $s_k$ and $x_k$.
In practice, $p(x|s)$ is usually specified according to the available \emph{a priori} knowledge coming from domain experts or a knowledge base.

We propose to adopt Gaussian mixture model (GMM) to formulate the \emph{a priori} knowledge that is used to specify $p(x|s)$. 
We select the GMM, because it is proved that any continuous pdf can be 
approximated by a mixture model \cite{zeevi1997density,bishop1995neural}.

For expository purposes, we present an example case of using GMM to model $p(x|s)$ in what follows. Suppose 
that we are concerned with a threat surveillance problem, and the situation space is $\mathcal{S}=\{\mbox{`danger'}, \mbox{`potential danger'}, \mbox{`safe'}\}$. The pdf $p(x|s)$ is modeled to be
\begin{equation}
p(x|s=\mbox{`danger'})=\sum_{i=1}^{M_d} \omega_{d,i} \mathcal{N}(x|X_{d,i},\Sigma_{d,i}),
\end{equation}
\begin{equation}
p(x|s=\mbox{`potential danger'})=\sum_{i=1}^{M_p} \omega_{p,i} \mathcal{N}(x|X_{p,i},\Sigma_{p,i}),
\end{equation}
\begin{equation}
p(x|s=\mbox{`safe'})=\sum_{i=1}^{M_{sa}} \omega_{sa,i} \mathcal{N}(x|X_{sa,i},\Sigma_{sa,i}),
\end{equation}
where $\mathcal{N}(\cdot|X,\Sigma)$ denotes a Gaussian pdf with mean $X$ and covariance $\Sigma$, $M$ denotes the number of mixing components in a mixture pdf, $\omega$ denotes the proportional mass of the mixing components, and the subscripts $d$, $p$ and $sa$ in $\omega$ respectively indicate the situations `danger', `potential danger' and `safe'. 
Assume that some domain or expert knowledge is available. Given such knowledge, 
the mixture parameters are specified correspondingly. It is noted that rather than manually specifying the model, its parameters
could be learnt from labelled historical data. Examples of learning a mixture
model to represent the state of moving targets are given in \cite{laxhammar2008anomaly,lane2012track}.

At this moment, all the details that is required to calculate the posterior, Equation (\ref{posterior_hmm}), has been completely presented, 
while another nontrivial issue about computation has to be considered if no analytic close-form solution to $p(y_k|s_k)$ as shown in 
Equation (\ref{likelihood_hmm}) is existent. We resort to the stochastic simulation techniques to approximate the integral in 
Equation (\ref{likelihood_hmm}). First we draw random samples, $\hat{x}_1, \hat{x}_2, \ldots, \hat{x}_N$, from $p(x|s_k)$. Assume that the sample 
size $N$ is large enough, the likelihood $p(y_k|s_k)$ can be approximated to be：
\begin{equation}\label{monte_carlo_hmm}
p(y_k|s_k)\approx\frac{1}{N}\sum_{i=1}^Np(y_k|\hat{x}_i).
\end{equation}
According to the large number theory, the accuracy of estimate improves as $N$ tends to infinity.

Note that the above inference process only produces an estimate of the posterior $p(s_k|y_{1:k})$, while provides no information about $x_k$, as $x$ 
is marginalized out in Equation (\ref{likelihood_hmm}). In the next subsection, we introduce a novel model, ESSM, based on which both the posterior 
of $x_k$ and that of $s_k$ can be estimated in a principled manner.
\subsection{Mixture based ESSM for SAW}
\begin{figure*}[htb]
\centerline{\includegraphics[angle=0,height=4cm,width=5cm]{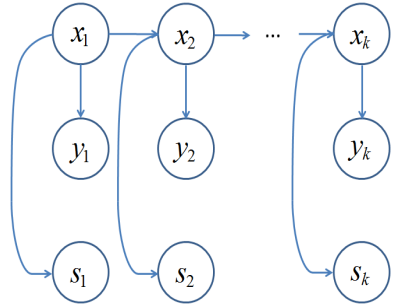}}
\caption{Extended State Space Model (ESSM) for SA}\label{ESSM}
\end{figure*}

Here we propose a SAW model, ESSM, which is an extension of the state space model that finds 
applications in tracking problems \cite{durbin2012time,arulampalam2002tutorial}. 
The ESSM model is graphically illustrated in Fig.\ref{ESSM}, where each arrow indicates a 
dependence and $k$ denotes the discrete time step, the same as for Fig.\ref{hmm}. 
The physical meaning of the hidden variable $x$ could be, e.g., the position and velocity 
elements of the moving targets of our interest in a surveillance region. 
The evolution of $x$ is specified by a Markov model $p(x_{k+1}|x_k)$. 
The situation $s$, is directly dependent on $x$ of the same time. 
The only observable variable $y$ denotes the sensor measurements. 
The same as $s$, $y$ is also dependent on $x$ of the same time. 
It is worthy to note that the dependence relationship between $s$ and $x$ and 
that between $y$ and $x$ is totally different. The former is problem specific 
and is determined by the expert or domain knowledge, while the latter is just 
determined by the sensor type. The same as in the Sec. \ref{sec:hmm}, we formulate 
the knowledge that is needed to model the dependence relationship between $s$ and $x$ 
based on the GMM.

Given the model structure of ESSM, the posterior $p(s_k|y_{1:k})$ can be calculated as follows
\begin{equation}\label{posterior_ESSM}
p(s_k|y_{1:k})=\int_{\mathcal{X}} p(x_k,s_k|y_{1:k})dx_k=\int_{\mathcal{X}} p(s_k|x_k)p(x_k|y_{1:k})dx_k,
\end{equation}
where
\begin{equation}
p(s_k|x_k)=\frac{p(s_k,x_k)}{p(x_k)}=\frac{p(x_k|s_k)p(s_k)}{\sum_{s\in\mathcal{S}} p(x_k|s)p(s)}.
\end{equation}
In this model, we assume that $p(s)$ is a uniform distribution, so we have
\begin{equation}\label{s_conditional_on_x}
p(s_k|x_k)\varpropto p(x_k|s_k),
\end{equation}
where $p(x|s)$ is modeled by a GMM, in the same way as presented in Sec. \ref{sec:hmm}.

The posterior pdf of $x$, namely $p(x_k|y_{1:k})$, is calculated in a sequential manner as follows. Given $p(x_{k-1}|y_{1:k-1})$, we have
\begin{equation}\label{state_filtering}
p(x_k|y_{1:k})=\frac{p(y_k|x_k)\int_{\mathcal{X}} p(x_{k-1}|y_{1:k-1})p(x_k|x_{k-1})dx_{k-1}}{p(y_k|y_{1:k-1})}.
\end{equation}
Suppose that both the state transition prior $p(x_k|x_{k-1})$ and the likelihood function $p(y_k|x_k)$ 
have been defined appropriately, then the state filtering algorithms, such as the Kalman filter 
and its variants or particle filtering (PF) methods can be used straightforwardly here to 
calculate Equation (\ref{state_filtering}).

Here we present a PF solution, since the PF can deal with more complex nonlinear and/or non-Gaussian cases.
To begin with, let $\{\hat{x}_k^i, w_k^i\}_{i=1}^N$ denote a random measure that approximates $p(x_k|y_{1:k})$, which means
\begin{equation}\label{monte_carlo_approx}
p(x_k|y_{1:k})\simeq \sum_{i=1}^N w_k^i\delta(x_k-\hat{x}_k^i),
\end{equation}
where $\delta(\cdot)$ denotes the Dirac delta function.

Assume that at time step $k-1$, a discrete weighted sample set $\{\hat{x}_{k-1}^i, w_{k-1}^i\}_{i=1}^N$, which
approximates $p(x_{k-1}|y_{1:k-1})$, is available, the task is to get a particle approximation for
$p(x_k|y_{1:k})$. Employing $p(x_k|x_{k-1})$ as the proposal distribution, the importance weights
can be determined based on the principle of importance sampling \cite{doucet2000sequential}. Specifically, given
$\hat{x}_{k-1}^i$, draw a random sample $\hat{x}_k^i$ from the state transition prior, and then calculate the importance weight
as follows
\begin{equation}
w_{k}^i=\frac{p(y_k|\hat{x}_k^i)}{\sum_{i=1}^N p(y_k|\hat{x}_k^i)}, i=1,2,\ldots,N.
\end{equation}
Then a Monte Carlo estimate to $p(x_k|y_{1:k})$ is available, as shown in Equation (\ref{monte_carlo_approx}).
A resampling procedure is often used to avoid particle divergence, see details in \cite{arulampalam2002tutorial}.
Here we present a simple way to implement the PF idea. See alternatives of PF implementations 
in \cite{arulampalam2002tutorial,doucet2009tutorial,van2000unscented}, for example. The convergence
properties of the PF methods for nonlinear non-Gaussian state filtering problems have been 
proved \cite{crisan2002survey,arulampalam2002tutorial,hu2008basic}.

Now we substitute $p(s_k|x_k)$ and $p(x_k|y_{1:k})$ in Equation (\ref{posterior_ESSM}) with Equations  (\ref{s_conditional_on_x}) 
and (\ref{monte_carlo_approx}), respectively, and then we obtain
\begin{equation}
p(s_k|y_{1:k})\varpropto\sum_{i=1}^N w_{k}^i p(\hat{x}_k^i|s_k).
\end{equation}
Because $\sum_{s_k\in\mathcal{S}}p(s_k|y_{1:k})=1$, we could get a particle approximation of the posterior $p(s_k|y_{1:k})$ as follows
\begin{equation}
p(s_k|y_{1:k})=\frac{\sum_{i=1}^N w_{k}^i p(\hat{x}_k^i|s_k)}{\sum_{s_k\in\mathcal{S}}\sum_{i=1}^Nw_{k}^i p(\hat{x}_k^i|s_k)}.
\end{equation}
\section{Simulation results}
For expository purposes, we introduce an application of the proposed methods in a toy example. 
We design a simulation experiment which is similar to the suspicious incoming smuggling vessel 
case presented in \cite{fischer2014modeling}. The objective here is to demonstrate that the proposed
mixture idea works. A comparative study of our method
with other related methods, for example, MEBN, is certainly interesting but has not yet been
performed and is not the intention here.

\begin{figure*}[!htb]
\centerline{\includegraphics[angle=0,height=12cm,width=15cm]{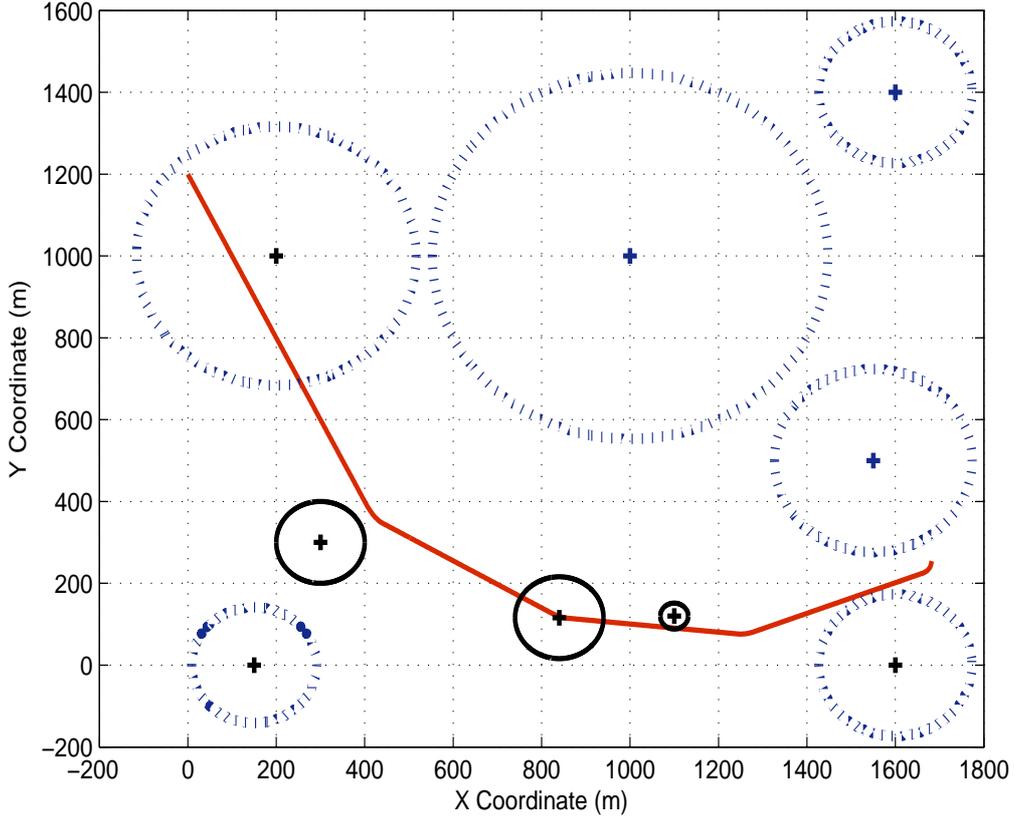}}
\caption{A graphical description of the simulation settings. The sensor is located at the origin. The target's trajectory is denoted by 
the solid lines. The circles drawn with solid and dot lines denote the standard error ellipses associated with the covariance matrices of the 
mixture pdf conditional on the `danger' situation and those on the `safe' situation, respectively.}\label{Traj}
\end{figure*}

The experiment is about a simulated scenario of threat surveillance. In this scenario, a target moving within a 
two dimensional surveillance area is monitored by a sensor, which generates noisy bearing and radial distance 
observations all the time.
Within this surveillance area, there are some sensitive regions. Once the target enters into such regions, it 
indicates that a danger will happen. The task it to design an algorithm that can replace the human operator by 
automatically percepting and predicting the appearance of the dangerous events in real time. Here the situation 
parameter space is $\mathcal{S}=\{\mbox{`danger'},\mbox{`potential danger'},\mbox{`safe'}\}$.
See Fig.\ref{Traj} for a graphical illustration of the experimental setting, where the solid line denotes the 
target's trajectory, and the circles denote the sensitive regions. The target begins moving from the upper left 
region of the surveillance area.

In this experimental case, the hidden variable $x$ represents a vector including the two dimensional position and velocity elements of the moving 
target, i.e.,
\begin{equation}
x_{k}=\left[\textbf{\mbox{x}}_k\quad \dot{\textbf{\mbox{x}}_k}\quad \textbf{\mbox{y}}_k\quad \dot{\textbf{\mbox{y}}_k}\right]^T,
\end{equation}
where $(\textbf{\mbox{x}},\textbf{\mbox{y}})$ and $(\dot{\textbf{\mbox{x}}},\dot{\textbf{\mbox{y}}})$ denote 
the two dimensional position and velocity respectively, and $A^T$ denotes the transposition of vector $A$.
The target's movement is characterized by a near constant velocity model as follows
\begin{equation}\label{state_trans}
\textbf{\mbox{x}}_{k+1}=\textbf{\mbox{f}}_{k}(\textbf{\mbox{x}}_{k},\textbf{\mbox{v}}_{k}),
\end{equation}
where
\begin{equation}\label{constant_velocity_model}
\textbf{\mbox{f}}_{k}(\textbf{\mbox{x}}_{k},\textbf{\mbox{v}}_{k})=\textbf{\mbox{F}}\textbf{\mbox{x}}_{k-1}+\textbf{\mbox{v}}_{k},
\end{equation}
$
\textbf{\mbox{F}}=\left[ \begin{array}{cc}
\textbf{\mbox{F}}_{\textbf{\mbox{s}}} & 0 \\
0 & \textbf{\mbox{F}}_{\textbf{\mbox{s}}}\end{array}\right],$
$
\textbf{\mbox{F}}_{\textbf{\mbox{s}}}=\left[\emph{}\begin{array}{cc}
1 & \mbox{T} \\
0 & 1\end{array}\right]
$, $\mbox{T}$ denotes the sampling period of the measurements, and $\textbf{\mbox{v}}$ is the process noise, 
which is zero-mean Gaussian distributed with covariance $
\textbf{\mbox{Q}}=\textbf{\mbox{B}}\left[ \begin{array}{cc}
10 & 0 \\
0 & 10\end{array}\right]\textbf{\mbox{B}}^T$, where 
\begin{equation}
\textbf{\mbox{B}}=\left[ \begin{array}{cc}
\mbox{T} & 0 \\
0 & \mbox{T}\\
\mbox{T}^2/2 & 0\\
0 & \mbox{T}^2/2
\end{array}\right].
\end{equation}

The distribution of the target state $x$ conditional on the situation parameter $s$, i.e., $p(x|s)$, is specified to be
\begin{equation}
p(x|s=\mbox{`danger'})=\sum_{i=1}^{M_d} \omega_{d,i} \mathcal{N}(x|X_{d,i},\Sigma_{d,i}),
\end{equation}
\begin{equation}
p(x|s=\mbox{`potential danger'})=\sum_{i=1}^{M_p} \omega_{p,i} \mathcal{N}(x|X_{p,i},\Sigma_{p,i}),
\end{equation}
\begin{equation}
p(x|s=\mbox{`safe'})=\sum_{i=1}^{M_{sa}} \omega_{sa,i} \mathcal{N}(x|X_{sa,i},\Sigma_{sa,i}).
\end{equation}
The number of mixing components in $p(x|s=\mbox{`danger'})$ is 3, see Fig.\ref{Traj} for a 
graphical description of this mixture pdf. The weights of the mixing components in this mixture 
pdf is fixed to be $1/3$ and the covariance matrix $\Sigma_{d,i}$ is diagonal.

The mixture pdf $p(x|s=\mbox{`potential danger'})$ is set to be the same as $p(x|s=\mbox{`danger'})$, except that the diagonal elements 
of $\Sigma_{p,i}$ is 10 times bigger than those of $\Sigma_{d,i}$.

The mixing components in $p(x|s=\mbox{`safe'})$ are all dispersive distributions over the surveillance regions 
excluding the danger and potential danger regions, see Fig.\ref{Traj} for a graphical description of the mixture pdf
conditional on the `safe' situation.

In the simulation, the sensor generates noisy measurements including the relative bearing $\theta$ and radial distance $r$ of the target, 
with respect to the sensor. The sensor noise is Gaussian distributed. The standard errors of the bearing and the radial distance 
measurements' distribution are set to be 0.1 degree and 50 meters, respectively.

\begin{figure*}[!htb]
\centerline{\includegraphics[angle=0,height=8cm,width=12cm]{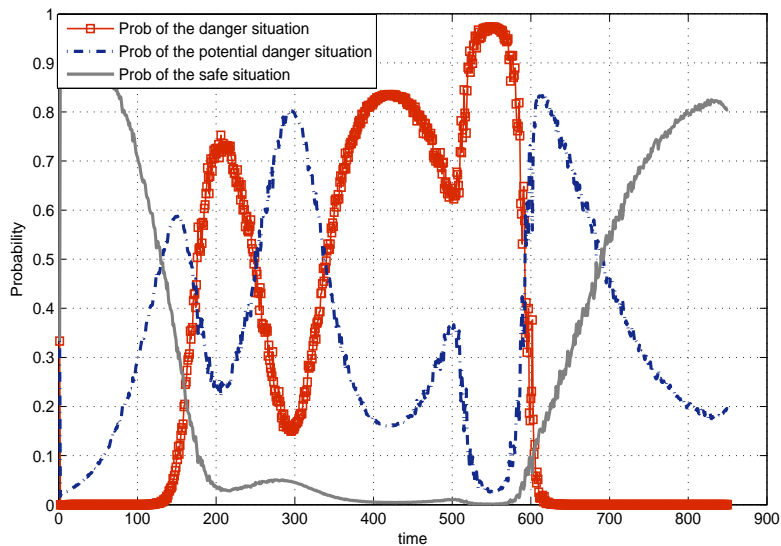}}
\caption{The real time situation awareness result by using the proposed HMM model.}\label{HMM_Situation_Prob}
\end{figure*}

First we apply the proposed HMM model to this scenario in order to test its effectiveness. The sample size $N$ in Equation (\ref{monte_carlo_hmm}) 
is set to be 10,000. The state transition process $p(s_{k+1}|s_k)$ is determined by the transition table as below
\begin{equation}\label{Trans_matrix}
\begin{array}{cccc}
~ & \mbox{`safe'} & \mbox{`potential danger'} & \mbox{`danger'}\\
\mbox{`safe'}& 0.9 & 0.1 & 0 \\
\mbox{`potential danger'} & 0.05 & 0.9 & 0.05 \\
\mbox{`danger'} & 0 & 0.1 & 0.9  \end{array}.
\end{equation}
One example run of the HMM based SAW method gives the real time estimate of the posterior, 
$p(s_k|y_1,\ldots,y_k)$, as shown in Fig.\ref{HMM_Situation_Prob}. We see that at the beginning, the posterior 
probabilities of the three situations is the same, then the posterior probability of $s=\mbox{'safe'}$ rises 
abruptly to be close to 1 and then falls off gradually. The above changes in the output of the approach reflects 
accurately the initial phase of the experiment when the target has just entered the surveillance region, moving 
closer to the first sensitive region. From Fig.\ref{HMM_Situation_Prob}, we see that the event $s=\mbox{`potential danger'}$ 
is most probable after the 130th time step and then the event $s=\mbox{`danger'}$ becomes most probable after 
about the 180th time step. This is totally consistent with the fact that the target moves closer to the first 
sensitive region and then enters it during that period. The similar analysis can be performed for the remaining 
processes of the target's movement, and it could be found that the output result of the SAW approach is consist 
with the truth. 

It should be noted that, in Fig.\ref{Traj} the target moves through the centre of `danger' 
region 2 and touches the edge of `danger' region 3, while the maximum probability of `danger' in
Fig.\ref{HMM_Situation_Prob} occurs for region 3. This unexpected observation indicates that the result yielded 
by the HMM model is not optimal. 

\begin{figure*}[!htb]
\centerline{\includegraphics[angle=0,height=8cm,width=12cm]{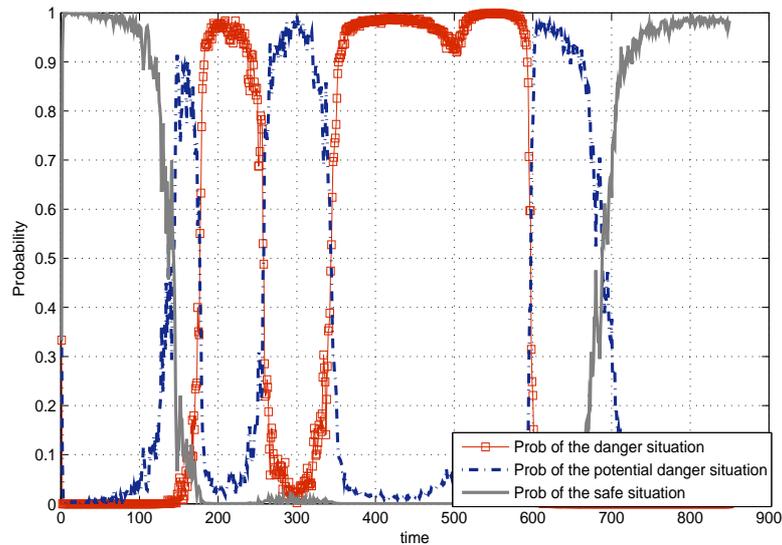}}
\caption{The real time situation awareness result by using the proposed ESSM model.}\label{ESSM_Situation_Prob}
\end{figure*}

\begin{figure*}[!htb]
\centerline{\includegraphics[angle=0,height=8cm,width=12cm]{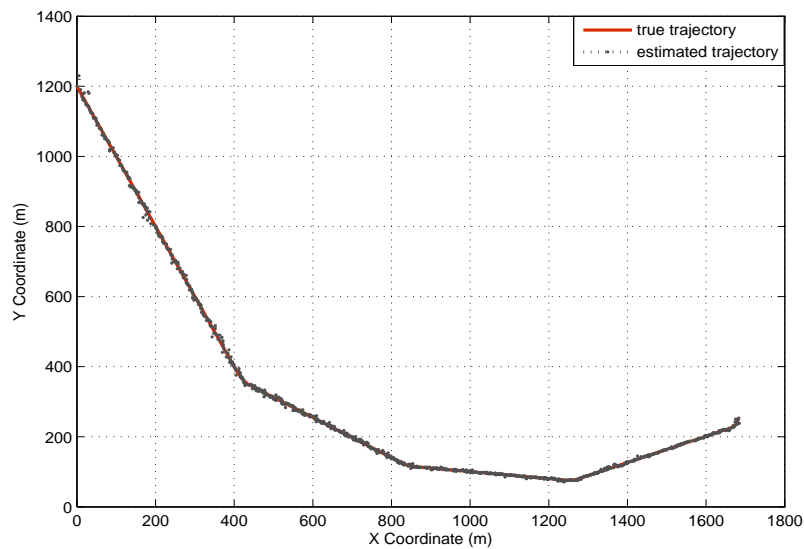}}
\caption{The target tracking result given by the PF based on the ESSM model.}\label{Track}
\end{figure*}

Next we run the ESSM model based SAW approach for this scenario to verify its effectiveness. 
The particle size used in PF is set to be 5000. The resulting SAW result is shown in 
Fig.\ref{ESSM_Situation_Prob}. 
Observe that, every time the target enters a sensitive region, the posterior probability of the `danger' 
situation approaches 1. 
So, in comparison with the HMM model, the ESSM model is shown to have advantage in leading to more credible
detections of the `danger' situation. For other phases of the target's movement process, the ESSM model always 
produce expected result that is consistent with the truth. 
Besides, the ESSM based approach can provide a byproduct, a real time estimate of the target state $x$. 
The real 
time estimate of $x_k$ outputted from an example run of the ESSM based approach is plotted in Fig.\ref{Track}. 
As is shown, the estimated target trajectory matches the truth very well.

Note that the result shown before is not intentionally selected. For both models, we have run the inference 
algorithm for many times, and the results are very similar as those shown above.
\section{Conclusions}
In this paper, we studied a statistical modeling approach to do SAW and proposed the idea of using mixture 
models to formulate the expert/domain knowledge that is required for building the SAW model. 
We presented two instantiation models, HMM and ESSM, whose implementation is assisted by a GMM based 
representation of the expert/domain knowledge. The efficiency of the proposed approach is testified by 
a toy simulation experiment. It is shown that, in comparison with HMM, the ESSM model has advantages in producing more reliable 
estimation of the situations. Utilization of the ESSM based approach also provides 
a byproduct, a real time estimate of the target state $x$, which is modeled as a pivot 
variable that connects the sensor measurement and the situation.

A promising future work consists of a further investigation of the mixture based 
approach for formulating the expert/domain knowledge in the context of SAW and a theoretical as well as 
empirical comparison of the proposed approach with the MEBN based methods.
\bibliographystyle{IEEEtran}
\bibliography{mybibfile}
\end{document}